\documentclass[11pt]{article}  %Do not change the point size.
\usepackage{menuproc}
\usepackage{epsfig}
\usepackage{amsmath,amssymb}
\begin{document}
%
%____________________________________________________________
%
%  Title, authors, institutions, and abstract
%----------------------------------------------------------------
%  Syntax:  \titlematter{title}{authors}{institutions}{abstract}
%----------------------------------------------------------------
%     If lines are too long, use linebreaks where convenient.
%     If all authors are from the same institution, omit raised letters.
%
\titlematter{$\pi$ and $\pi\pi$ Decays of Excited $D$ Mesons}%
{T.A. L\"ahde$^a$ and D.O. Riska$^a$}%
{$^a$Helsinki Institute of Physics,\\
     University of Helsinki, PL 64 Helsinki, Finland}%
{The $\pi$ and $\pi\pi$ decay widths of the excited charm mesons are 
calculated using a Hamiltonian model within the framework of the 
covariant Blankenbecler-Sugar equation. The pion-light constituent quark 
coupling is described by the chiral pseudovector Lagrangian.} % %
%____________________________________________________________
%  Start article here:

%%%%%%%%%%%%%%%%%%%%%%%%%%%%%%%%%%%%%%%%%%%%%%%%%%%%%%%%%%%%%%%%%%%%%%%%%%%%%%%%%%
\section{Introduction}

The pionic decay widths of the excited charm mesons ($D$ mesons) are 
interesting observables, since they depend straightforwardly on the 
coupling of pions to light constituent quarks. The $D$ mesons consist of 
one light ($u,d$) quark and a heavy charm ($c$) antiquark, of which it is 
only the light constituent quark that couples to pions. The coupling of light 
constituent quarks to pions may be described by the chiral 
model~\cite{Holstein}, which includes the pseudovector Lagrangian and, for 
$\pi\pi$ decay, also a Weinberg-Tomozawa term. 

In order to predict the decay widths of the 
excited $D$ meson states, a model for the radial 
wavefunctions is needed. Here the interaction between the quarks is 
modeled as the sum of a screened one-gluon exchange (OGE) and a scalar 
linear confining interaction. The wavefunctions are obtained as 
solutions of the covariant Blankenbecler-Sugar equation~\cite{M1dec}. 
These are then used together with the chiral Lagrangian to obtain 
predictions for the $\pi$~\cite{Pidec} and $\pi\pi$~\cite{2pidec} decays 
of the excited $D$ mesons. 

\begin{figure}[h]
\parbox{.55\textwidth}{\epsfig{file= 
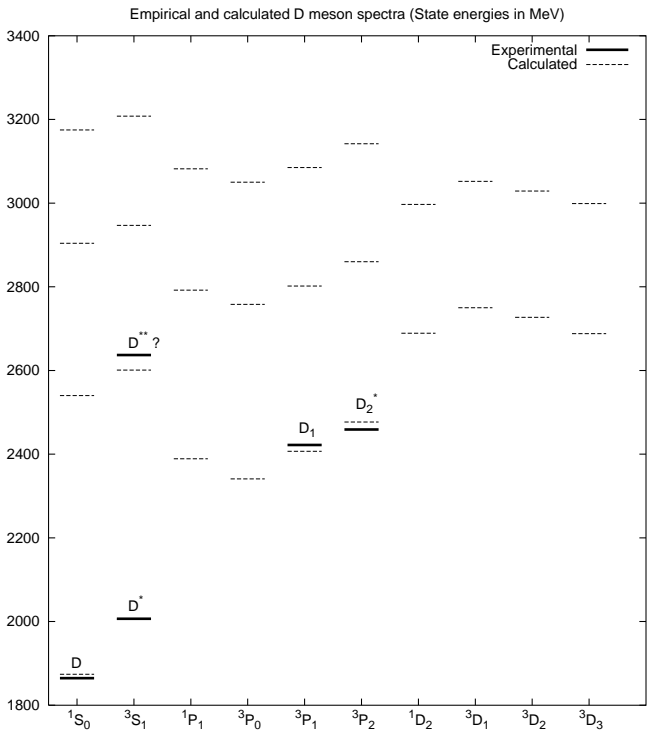,width=.55\textwidth,silent=,clip=}}
\hfill
\parbox{.4\textwidth}{\caption{Empirical and calculated spectra of the $D$ 
meson from ref.~\cite{Pidec}. The $D^*$ is an $S$-wave spin-triplet state 
which lies almost 
exactly at threshold for decay to $D\pi$. The decay widths for $\pi$ decay 
of the $D^*$ are predicted here along with those of the four $L=1$ states, 
which can decay to both $D^*\pi$ and $D\pi$. These states can also decay 
to $D^*\pi\pi$ and $D\pi\pi$. Note that empirical data is only available 
for the spin triplet states with $L=1$ and total angular momentum $J=1$ 
and $J=2$.}}
\end{figure}

\newpage

%%%%%%%%%%%%%%%%%%%%%%%%%%%%%%%%%%%%%%%%%%%%%%%%%%%%%%%%%%%%%%%%%%%%%%%%%%%%%%%%%%

\section{Single pion decay}

The chiral Lagrangian describing the coupling between pions and light 
constituent quarks may be written as~\cite{Pidec}
\begin{equation}
 {\cal L}=i{\frac{g_A^q}{2f_\pi}}\bar\psi_q\,\gamma_5\gamma_\mu
\,\partial_\mu\,\vec\phi_\pi\cdot\vec \tau\,\psi_q, \label{L1}
\end{equation}
where $g_A^q$ is the pion-quark axial coupling constant and $f_\pi$ is the 
pion decay constant. The Lagrangian~(\ref{L1}) gives rise to both axial 
current and charge single-quark amplitudes which may be obtained as
\begin{eqnarray}
T_P^{\mathrm{1q}}\!\!\!\!&=&\!\!\!\!-i\frac{g^q_A}{2f_\pi}
\sqrt{\frac{(E'+m_{\bar q})(E+m_{\bar q})}{4EE'}}
\left(1-\frac{P^2-k^2/4}{3(E'+m_{\bar q})(E+m_{\bar q})}\right) 
\vec\sigma^q\cdot\vec k\,\tau_\pi, \\
T_S^{\mathrm{1q}}\!\!\!\!&=&\!\!\!\! i\frac{g^q_A}{2f_\pi}
\frac{2m_{\bar q}+E+E'}{\sqrt{4EE'(E+m_{\bar q})(E'+m_{\bar q})}}
\:\omega_\pi\,\vec \sigma^q\cdot
\left(\frac{\vec p\,'+\vec p}{2}\right)\,\tau_\pi.
\end{eqnarray}
Here the relativistic momentum-dependent factors arise from the spinors of 
the light constituent quark. Because of the small mass (450 MeV) of the 
light constituent quark, these factors should not be dropped, as otherwise 
large overestimates will result. In the above expression, $\vec k$ denotes 
the momentum of the emitted pion. The quark momentum operator $\vec P$ is 
defined as ${(\vec p\,'+\vec p\,)/2}$. In addition to the above 
single-quark amplitudes, two-quark amplitudes involving excitation of 
intermediate negative energy quarks by the quark-antiquark interaction 
have been considered. Especially for the $\pi$ decays of the $L=1$ states, 
they are shown to have a large effect.

Using the standard phase-space expressions and wavefunctions for the 
initial and final $D$ meson states, the $\pi$ decay widths of these states 
may be predicted. The predictions for the decays of the $D^*$ agree well 
with the recent CLEO measurement~\cite{CLEO} of 
the width of the $D^{*\pm}$, which is reported as $96 \pm 4 \pm 22$ keV.

\begin{table}[h]
\parbox{.55\textwidth}{
\begin{center}
\begin{tabular}{c|c|c}
Decay & \quad $g_A^q = 0.87$ \quad & \quad $g_A^q = 1.0$ \quad \\
\hline\hline && \\
$D^{*\pm} \rightarrow D^{\pm}\pi^0$ & \quad 29 keV \quad & \quad 38 keV
\quad \\ % && \\
$D^{*\pm} \rightarrow D^0\pi^\pm$   & \quad 64 keV \quad & \quad 84 keV
\quad \\ % && \\
$D^{*0} \rightarrow D^0\pi^0$       & \quad 41 keV \quad & \quad 54 keV
\quad
\end{tabular}
\end{center}}
\hfill
\parbox{.4\textwidth}{\caption{Numerical results for the $D^*$ widths 
using $g_A^q = 0.87$ and $g_A^q = 1.0$. The CLEO measurement concerns 
the sum of the two decay modes of the $D^{*\pm}$. As of this time, no 
constraining data exists for the pion decay of the neutral $D^*$ meson.}} 
\end{table}

Predictions for the pion decay widths of the $L=1$ $D$ meson states have 
also been made in ref.~\cite{Pidec}. In case of $S$-wave decay of these 
states, it is shown that in order to avoid large overestimates it is 
important to include also two-quark exchange charge amplitudes associated 
with excitation of intermediate negative energy quarks by the 
quark-antiquark interaction. Generally, the predicted widths of the spin 
triplet states fall somewhat below the empirical values, which suggests 
that $\pi\pi$ decay may play a role.

%%%%%%%%%%%%%%%%%%%%%%%%%%%%%%%%%%%%%%%%%%%%%%%%%%%%%%%%%%%%%%%%%%%%%%%%%%%%%%%%%%

\section{Two-pion decay}

The Lagrangian~(\ref{L1}) gives rise to Born terms describing $\pi\pi$ 
decay of excited $D$ mesons. The model is completed by adding the 
Weinberg-Tomozawa Lagrangian
\begin{equation}
{\cal L}_{\mathrm{WT}} =
-\frac{i}{4f_{\pi}^2}\bar\psi_q\,\gamma_{\mu}\,
\vec\tau\cdot\vec\phi_{\pi}\times\partial_{\mu}
\vec\phi_{\pi}\,\psi_q. \label{L2}
\end{equation}
Together, the Lagrangians~(\ref{L1},\ref{L2}) give rise to amplitudes for 
$\pi\pi$ decay, which are usually expressed in the form $T = 
\delta_{ab}T^+ + \frac{1}{2}[\tau_b,\tau_a] T^-,$ where $a,b$ are isospin 
indices, and the amplitudes $T^\pm$ are given as $T^\pm = \bar u(p') 
\left(A^\pm - i\gamma\cdot Q B^\pm\right) u(p).$ Here $Q$ is defined as 
the combination $(k_b - k_a)/2$ of the pion four-momenta. The resulting 
expressions for the sub-amplitudes $A,B$ are obtained as
\begin{eqnarray}
A^+ &=& \left(\frac{g_A^q}{2f_{\pi}}\right)^2 4m_q, \\
A^- &=& 0, \\
B^+ &=& -\left(\frac{g_A^q}{2f_{\pi}}\right)^2 4m_q^2
\left[\frac{1}{s - m_q^2} - \frac{1}{u - m_q^2}\right], \\
B^- &=& -\left(\frac{g_A^q}{2f_{\pi}}\right)^2
\left(2 + 4m_q^2\left[\frac{1}{s - m_q^2} + \frac{1}{u - m_q^2}\right]
\right) + \frac{1}{2f_\pi^2}.
\end{eqnarray}
Computation of the two-pion decay widths of the excited $D$ mesons using 
the above amplitudes together with the appropriate phase space expressions 
is shown in ref.~\cite{2pidec} to lead to a significant increase of the 
total widths of the $L=1$ $D$ meson states, which is also expected from 
comparison with the analogous strange mesons~\cite{PDG}.

\begin{table}[h]
\parbox{.55\textwidth}{
\begin{center}
\begin{tabular}{c|c|c|c|c}
$D$ state & $\pi$ width & $\pi\pi$ width & Total &  Exp \\
\hline\hline &&&& \\
$D_2^*$ & 15.7 & 3.05 & 18.8 &
25$^{+8}_{-7}$ \\ &&&& \\
$D_1$   & 13.6 & 1.34 & 14.9 &
$18.9_{-3.5}^{+4.6}$ \\ &&&& \\
$D_0^*$ & 27.7 & $\sim$ 0.1 & 27.8 & -- \\ &&& \\
$D_1^*$ & 13.2 & 8.62 & 21.8 & -- \\ &&& \\
\end{tabular}
\end{center}}
\hfill
\parbox{.4\textwidth}{\caption{Summary of results for $\pi$ and $\pi\pi$ 
decay of the $L=1$ $D$ mesons in MeV, for $g_A^q = 1.0$. Empirical data 
is only available for the spin triplet $D_2^*$ and $D_1$ states. The 
large variations in the $\pi\pi$ widths are due to the large differences 
in phase space.}} 
\end{table}
In this context it is to be noted that the widths of the $D_0^*$ and the 
$D_1^*$ are here predicted to be an order of magnitude smaller than 
previously thought, because of the inclusion of negative energy 
components into the decay amplitudes; A similar effect has been noted in 
the calculation of ref.~\cite{Goity}. However, the results are strongly 
dependent on both the assumed spin-orbit structure of the $L=1$ states and 
the composition and coupling structure of the quark-antiquark interaction, 
and should therefore be viewed as suggestive rather than as definite 
quantitative predictions.

%%%%%%%%%%%%%%%%%%%%%%%%%%%%%%%%%%%%%%%%%%%%%%%%%%%%%%%%%%%%%%%%%%%%%%%%%%%%%%%%%%
%____________________________________________________________
%  Start references here:

\end{document}